\documentclass[preprint,12pt]{elsarticle}
\usepackage{graphicx,subfigure,epsfig}
\usepackage{amssymb}
\usepackage{epstopdf}
\usepackage{epsfig}
\usepackage{color}
\usepackage{soul}
\usepackage{float}
\usepackage{amsmath}
\usepackage{amssymb}
\usepackage{ifthen}
\usepackage{alltt}
\usepackage{fancyhdr}
\usepackage{verbatim}
\usepackage{moreverb}
\usepackage{colortbl}
\usepackage{multirow}
\usepackage{bigstrut}
\usepackage{wrapfig}
\usepackage{enumerate}
\usepackage{rotating}

\def\e{\begin{equation}}
\def\f{\end{equation}}
\def\=#1{\overline{\overline #1}}

\def\-#1{{\boldsymbol{#1}}}
\def\va{\varepsilon}
\def\o{\omega}

\begin{document}
\title{A System Analysis of Micron-Gap Thermophotovoltaic Systems Enhanced by Nanowires}

\author{Mohammad~Sajjad~Mirmoosa and Constantin~Simovski}

\address{Department of Radio Science and Engineering, School of Electrical Engineering, Aalto University, P.O. Box 13000, FI-00076 Aalto, Finland\footnote{e-mail address: mohammad.mirmoosa@aalto.fi}}

\begin{abstract}
We introduce new micron-gap thermophotovoltaic systems enhanced by tungsten nanowires. We theoretically show that
these systems allow the frequency-selective super-Planckian spectrum of radiative heat transfer that promises a
very efficient generation of electricity. Our system analysis covers practical aspects such as output power per unit area and efficiency of the tap water cooling.
\end{abstract}

\maketitle



\section{Introduction}

\begin{figure}[htbp]
  \centering
  \includegraphics[width=7cm]{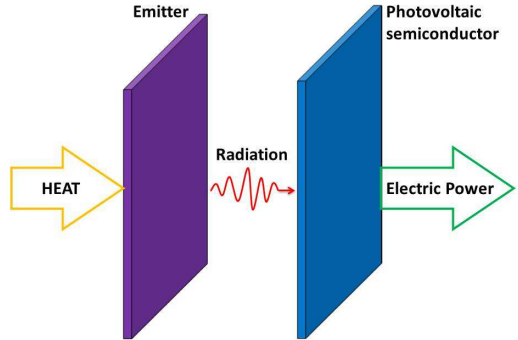}
\caption{Principle of a TPV system.}
 \label{TPV1}
\end{figure}

As an estimation, up to 50{\%} of the energy involved in the industrial processes finally delivers as waste heat.
Mankind needs to exploit this energy to generate electric power. It can be done indirectly, e.g. via vapor (Stirling's) machinery, or directly. Several methods
of the heat-electricity conversion are known: thermoelectric, pyroelectric, thermophotovoltaic, and thermophotogalvanic.
Unlike Stirling's machines such direct generators do not have moving parts and no permanent technical service is needed.
Thermophotovoltaic (TPV) conversion has attracted significant attention from the research community in the last decade \cite{B} due to
its potentially highest efficiency. TPV systems are based on the photovoltaic (PV) effect manifested by the photocurrent. The PV cell absorbs the thermal radiation
produced by the rear side of emitter whose front side is connected to the heat source, e.g. flame, see in Fig. \ref{TPV1}.
For high temperatures of the emitter,
corresponding to the near-infrared radiation (NIR), the TPV efficiency is higher compared to
other known mechanisms of the direct heat-electricity conversion \cite{Bauer}. Not only wasted heat can be converted, using TPV devices.
TPV generators with combustion cameras are also prospective for the domestic use \cite{B,Bauer}.

In spite of striking advantages of TPV devices as electric generators, a key problem of these systems is a disappointing gap between the maximally achievable and practically achieved operation parameters. For given temperatures of the emitter and the PV panel the electric power output per unit area (p.u.a.) of any known TPV system is much below its theoretically possible maximum \cite{C}. One of main reasons of this situation is non-advantageous, extremely broad spectrum of emitted radiation. It is commonly (and wrongly \cite{Rytov}) adopted that the maximal thermal radiation in all possible situations is radiation of a black body to free space. Therefore, the emitters mimicking the black body are often considered as best ones. The spectrum of its radiation has relative bandwidth (BW) which, defined on the 10\% level, exceeds 500\% even for temperature as high as $2000^{\circ}$K. The operational band of a PV semiconductor is much narrower -- its BW is nearly 100\%. Therefore, a large amount of thermal radiation in a TPV system is unusable \cite{D}. The radiation at frequencies below that of the semiconductor bandgap is completely harmful. At these frequencies, all thermal photons transmitted to the PV medium are dissipated and heat the PV panel destroying its operation. The PV operational band is narrower than the upper half of the thermal radiation spectrum (that above the bandgap frequency). Thus, the frequencies of thermal radiation which are twofold and greater than the bandgap frequency are also fully harmful.
The harmful action of the unusable spectrum is prevented in TPV systems with optical filters. These filters, roughly speaking, transmit only a useful part of the spectrum removing the reflection in this band. The harmful radiation is reflected. As a rule, these filters represent a multi-layer structure of transparent dielectrics \cite{B,Bauer}. It is clear that filtering though allows a TPV system to operate does not solve the problem of losses.

To avoid these losses is possible by squeezing the radiation spectrum compared to that of a black body. This regime may be offered by a frequency-selective emitter. Its radiation spectrum in the ideal case should mimic that of a black body within the PV operational band and vanish beyond it. Using frequency-selective emitters, one achieves high values of the TPV efficiency -- up to 20\% (see e.g. in \cite{S1}). Such advanced emitters may be based on photonic crystals or metamaterials. Especially high TPV efficiency -- up to 40\% -- is theoretically achieved for solar TPV systems, where advanced emitters also serve perfect absorbers of the sunlight \cite{S2,S3,S4}. However, the total radiation of any known advanced emitters over the PV operational band is noticeably lower than the Planckian limit. Practically, the spectrum of a metamaterial thermal emitter attains the black-body spectrum at an only resonant frequency. Also, beyond the operational band the thermal radiation of these emitters is not negligible. Therefore, the electric power output p.u.a. in these systems is not very high. For emitters with temperatures $T^{(1)}=2000^{\circ}$K this value in theoretical estimations \cite{S2,S3,S4} corresponds to 1.5--2 $\rm W/cm^2$ that is only twice as higher as that achieved in best available TPV systems operating at the same temperature \cite{Mauk}.

\section{Micron-gap TPV systems enhanced by hyperbolic metamaterials}

Recently, advanced TPV systems were introduced: so-called near-field systems comprising a nanogap between the PV medium and the emitter (see e.g. in \cite{NF}) and micron-gap ones fabricated by stacking the PV panel and the emitter separated by micron (or slightly submicron) spacers (see e.g. in \cite{M1,M2}). They possess so-called super-Planckian (SP) radiative heat transfer (RHT).
In a micron-gap TPV system the emitter transfers to the PV panel more power p.u.a. than the black body of the same temperature may do. The gain compared to a black-body emitter may be twofold
and occurs due to the so-called photon tunneling effect \cite{Bauer,D,NF,E}. In a near-field TPV system the photon tunneling is stronger and the gain compared to the black body may attain 3--4 orders of magnitude (see e.g. in \cite{Bauer,NF}). Unfortunately, the concept of a TPV filter is very difficult to implement for these advanced TPV systems: the RHT decays across a filter and the SP effect is lost. Therefore, in spite of high radiation fluxes, higher electric output compared to conventional TPV systems has not been claimed for these advanced devices. To avoid the harmful heating by unusable radiation, near-field TPV systems are only applied in the range of low, e.g. room, temperatures. Their practical purpose is temperature sensing \cite{Bauer,C,NF}.

\begin{figure}[htbp]
  \centering
  \includegraphics[width=10cm]{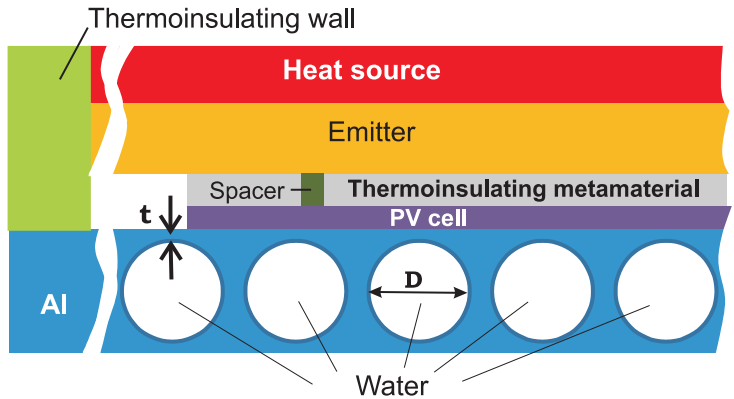}
\caption{A micron-gap TPV system enhanced by a metamaterial.}
 \label{TPV2}
\end{figure}

An actual target for specialists in TPV systems is to develop a frequency-selective emitter compatible with the concept of a near-field or a micron-gap TPV system. More exactly (since the emitter characteristics cannot be determined separately in such devices), one has to create TPV systems which offer the frequency-selective SP RHT. In the PV operational band this RHT should be above the Planckian limit and beyond this range -- sufficiently weak. Recently, some papers appeared on such a frequency-selective SP RHT in a near-field TPV system (e.g. \cite{Biehs}). However, we concentrate on micron-gap TPV systems. The last ones are more suitable for the generation of electricity because (unlike near-field TPV systems) they can be implemented on macroscopic area -- as large as  few $\rm cm^2$ -- whereas the gap between the emitting and cold (PV) surfaces can be as tiny as 500 nm \cite{M2}. The parallelism of these surfaces over a so substantial area is offered by sparsely located hollow quartz spacers \cite{M2} and is adjusted by springs \cite{patent}.

In micron-gap TPV systems, the strongly SP RHT (exceeding the black-body limit by one order of magnitude or more) is theoretically achievable using hyperbolic metamaterials (HMMs) filling the micron gap between the hot and cold media. A layer of HMM (and therefore multi-layer structures of HMMs) supports SP RHT beyond the photon tunneling \cite{Biehs,ThermoPRB,Hyper1,Hyper2}. It is possible to avoid the harmful contact between the hot and cold parts performing the HMM as it was suggested in \cite{ThermoPRB}. Then this metamaterial is thermoinsulating. In \cite{OE} it was theoretically shown that the properly designed HMM of metal nanowires offers the frequency-selective RHT needed for the generator applications. This results opens the door to micron-gap TPV systems with very high electric output.

\begin{figure}[htbp]
  \centering
  \includegraphics[width=10cm]{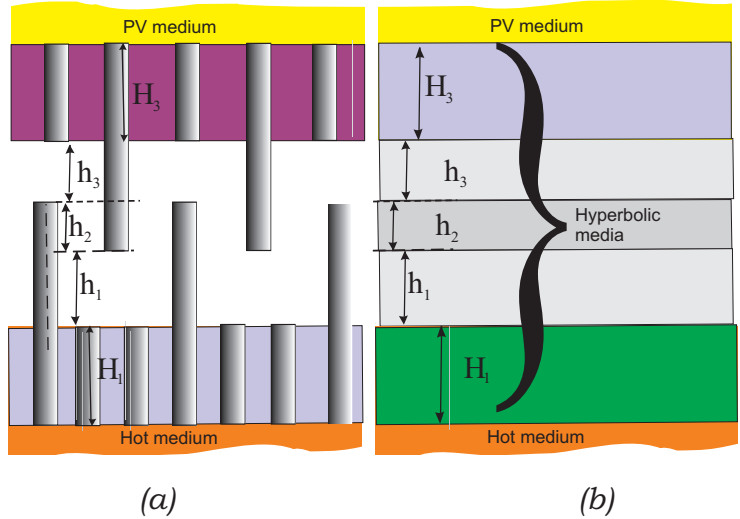}
\caption{(a) -- A seven-layer metamaterial structure with free-standing nanowires that offers the frequency-selective SP RHT
across five effective submicron layers. (b) -- An effective-medium model of this structure.}
 \label{TPV11}
\end{figure}

The TPV system can be implemented as it is shown in Fig. \ref{TPV2}. To respect the thermal balance in the steady regime, the heat generated in the PV panel needs to be evacuated.
The cooling system for practical electricity generation should be simple and passive (no energy supply). It may be a standard cooling system using the tap water flow.
Such the cooling structure is performed as a hollowed Al plate filled with flowing tap water. It is a standard cooling structure for TPV systems operating on the temperatures $1500^{\circ}{\rm{K}}<T^{(1)}<2500^{\circ}$K \cite{Book}. In Fig. \ref{TPV2}, we have shown a micron (or submicron) spacers, however, we do not consider the conductive heat transfer through them. Experiments have shown that this effect is negligible \cite{Bauer}. In accordance to \cite{Book} the conductive heat transfer through the insulating walls of the whole system may be also negligible. The operation of the system is practically determined by the RHT from the emitter to the PV cell though the thermo-insulating metamaterial. The implementation of the metamaterial \cite{OE} is shown in Fig. \ref{TPV11}. The left panel depicts two arrays of nanowires (hot and cold ones) partially free standing in the micron vacuum gap.
The right panel illustrates the effective-medium model (EMM) of the structure which explains the effect in terms of layered HMMs. The very high frequency selectivity of RHT, much better than that achievable for dielectric multilayers, results here from the strong optical contrast between involved effective layers \cite{OE}.

Our work \cite{OE} was only our first step to the creation of advanced TPV devices. Unfortunately, the scheme depicted in Fig. \ref{TPV11} have not stood further systematic analysis. In our work it was wrongly assumed that the melting point of 50 nm-thick gold nanowires is the same as that of bulk gold. In fact, it is not so, and a recent study \cite{MG} has shown that such gold nanowires are molten at noticeably lower temperature than $T^{(1)}=1300^{\circ}$K assumed to be the emitter temperature in \cite{OE}. So, gold nanowires should be replaced by refractory ones, e.g. tungsten ones, which can stand up to $T^{(1)}=2000-2100^{\circ}$K \cite{MT}. This replacement significantly changes the operation of effective HMM layers and whole design should be revised.

Second, in \cite{OE} we have not considered the heating of the PV medium. Even the heating caused by useful radiation is important, since the PV conversion of photons within the operational band is far from the perfect one. Not only the unusable part of the radiation spectrum (it was reduced to the safe values), even its useful part may violate the working temperature of the PV panel.  We will see below that for the interdigital arrangement of nanowires the standard water cooling system does not allow the thermal balance. In other words, the geometry depicted in Fig. \ref{TPV11} offers an excessive RHT in the useful frequency range. Thus, a modification of the geometry shown in Fig. \ref{TPV11} is needed also in order to reduce the RHT in the useful band to the reasonable level. Below we suggest two new prospective structures and present their systematic study.

\section{Main ideas and methods of study}

\begin{figure*}[!t]
    \centering
    \subfigure[$$]
    {
        \includegraphics[width=6.4cm]{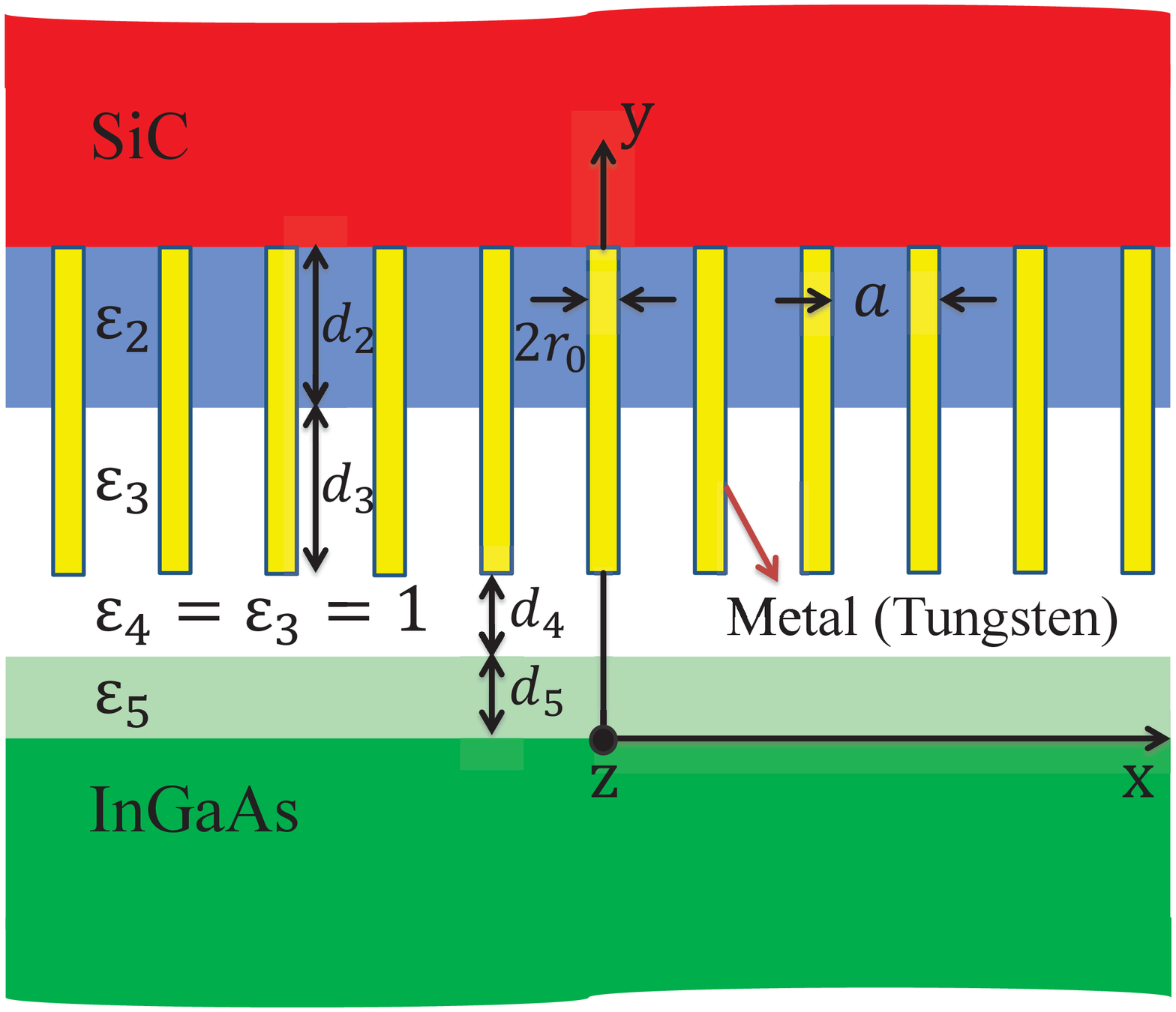}
        \label{fig:d_1}
    }
    \subfigure[$$]
    {
        \includegraphics[width=6.4cm]{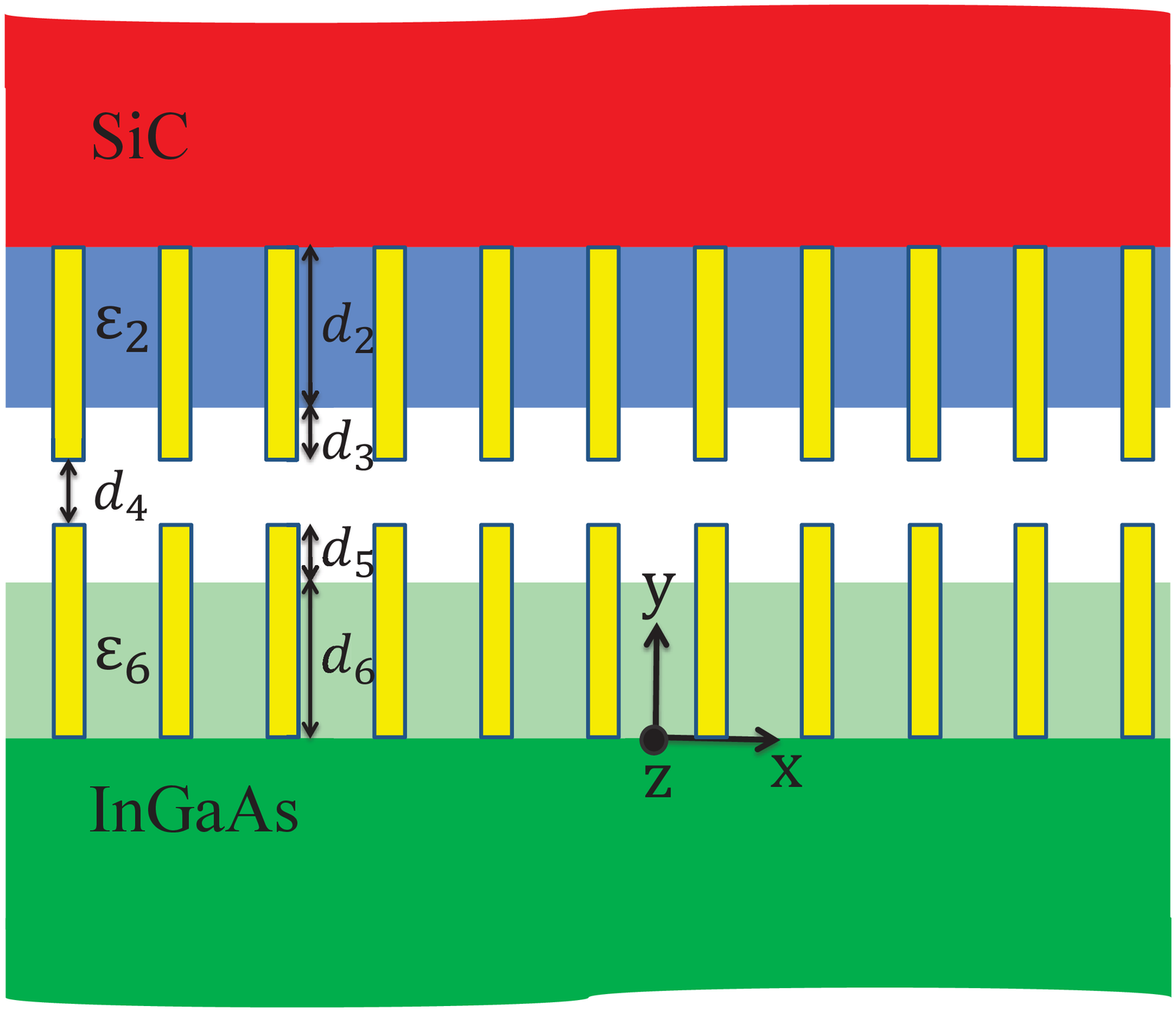}
        \label{fig:d_2}
   }
    \caption{Schematic view of the metamaterial structures under study. (a) -- An effectively 6-layer structure with free-standing nanowires grown in medium 2 and bonding nano-film 5. (b) -- An effectively 7-layer structure with hot and cold nanowitres.}
    \label{fig:d}
\end{figure*}

Two new design solutions of a prospective micron-gap TPV system with frequency-selective SP RHT are illustrated by Fig. \ref{fig:d}. Nanoiwres in both structures should be aligned. However, their periodicity is not required, only the correct fraction is important. Also, in the second structure hot and cold nanowires are not obviously located collinearly. Since both these structures comprise effective HMM multilayers and a vacuum nanogap $d_4$, their frequency selectivity should not be worse than that of the structure depicted in Fig. \ref{TPV11}. A certain reduction of the useful RHT may hold due to the presence of the vacuum nanogap. In our first structure, Fig. \ref{fig:d_1}, this gap is in between the hot nanowires and the nanofilm 5, and in our second structure, Fig. \ref{fig:d_2}, it is in between the hot and cold nanowires. This reduction is not a drawback of the new structures. On the contrary, it makes the TPV system compatible with the standard cooling. Another advantage compared to our previous design is, perhaps, the most important one. Both new structures are simpler for fabrication than the interdigital structure of nanowires. For the last one it would be very difficult to avoid the contacts between hot and cold nanowires. In accord to our estimations, if at least 30\% of top and bottom nanowires mutually touch, the conductive heat transfer through them becomes of the same order of magnitude as the RHT. In this condition the system cannot operate as an electric generator. The problem of possible contacts between hot and cold parts of the system is avoided if the gap $d_4$ is not smaller than the practical deviation of the nanowire length $\delta$, dictated by fabrication tolerances. Then the conductive heat transfer through the HMM becomes negligible.

In our work \cite{OE} the PV material was germanium, material having the bandgap energy 0.66 eV. We have revised this choice. In recent TPV systems, different kinds of semiconductors have been used which belong either to the group IV (Si, Ge, SiGe, etc.) or to the group III-V (GaSb, InGaAs, InGaAsSb, InPAsSb, etc.). The group IV corresponds to high-frequency devices which exploit the so-called inter-band NIR. Though the PV spectral response of these materials is high, such TPV systems are characterized by a large portion of unusable radiation.
Only a rather small portion of the radiated spectrum (for realistic temperatures of the emitter $T^{(1)}<3000^{\circ}$K) may correspond to these frequencies.
The deal between the maximal spectral response and the maximal useful portion of the radiated spectrum for this group of semiconductors results in the choice of germanium, as in \cite{OE}. However, applying the group III-V the same spectral response (and therefore the same quantum efficiency) is achievable together with lower bandgap energy and hence wider useful spectrum. For $\rm{In_xGa_{1-x}As}$ one obtains $E_{\rm{g}}$ from 0.36 to 1.42 eV varying the content of indium x from $1$ to $0$. For example, a micron-thick p-n layer of $\rm{In_{0.68}Ga_{0.32}As}$ has been grown on an InP substrate. This material has the bandgap $E_{\rm{g}}=0.6$ eV \cite{Bauer}. In accordance to \cite{Bauer} efficiency parameters of this PV cell working on the temperature 25$^{\circ}$C are as follows: $\eta_{\rm{OC}}=53.9{\%}$ (open-circuit or voltage factor), $\eta_{\rm{QE}}=75.3{\%}$ (mean quantum efficiency), and $\eta_{\rm{FF}}=71.5{\%}$ (fill factor). Furthermore, 0.55 eV is achievable with ${\rm{x}}=0.6$ \cite{Bauer}. The complex refractive index of InGaAs for both these modifications is practically the same and its frequency dependence can be found in~\cite{InGaAs}. We assume that two corresponding PV cells may have the same efficiency parameters. The relative BW of the PV operational band for such PV cells is equal 100-110\%. Using structures shown in Fig. \ref{fig:d} the spectrum of RHT can be fit to this band for both modifications of InGaAs that explains our choice.

Though our initial speculations on HMMs enhancing the RHT were done for semi-transparent emitters, we may apply the same schemes with a metallic emitter. Thus, in our last simulations we have replaced the silicon carbide emitter by a tungsten one. The purpose of this replacement was twofold. First, it is easier to grow W nanowires on a W substrate than inside a thin dielectric host on top of a SiC substrate as it is implied by Fig. \ref{fig:d}. Second, the optical contrast between W and the HMM layers is stronger than that in the case of the SiC emitter. This enhanced contrast may result in further squeezing the RHT spectrum that enhances the PV efficiency. Simulations have confirmed our expectations.

To calculate the RHT across the HMM structure we apply the original method which is developed in \cite{Circuit Model}. This method is called the equivalent circuit method for calculating the RHT. It is based on the expansion of the RHT into frequency and spatial harmonics and construction of the impedance matrix for each spatial harmonic in every (perhaps, structured) layer of the system. This model assumes a steady regime characterized by a given temperature distribution over the layers. The RHT is calculated as a sum of radiative heat fluxes generated by every hot layer and transferred to the PV layer. In principle, the method \cite{Circuit Model} allows the impedance matrix to be built for an arbitrary periodic structure of any thickness, such as an array of parallel nanowires. However, an EMM is a strong simplification of this problem that offers a huge economy of the computational time. Therefore, in order to perform a numeric optimization of the structures shown in Fig. \ref{fig:d}, it is very important to check how accurate this model of a HMM layer is when calculating the RHT. Fortunately, the EMM is an adequate and efficient tool for the analysis of RHT in multilayers filled with metal nanowires \cite{file}. Moreover, in \cite{file} we have shown that for multilayers with tungsten nanowires the simplest variant of the EMM -- a quasi-static EMM (QEMM) -- ensures the good accuracy for the RHT spectrum. Therefore, we combine the QEMM and the equivalent circuit method.

Further, we calculate the radiative heating of the PV cell p.u.a. through the RHT taking into account the overall PV efficiency of the chosen PV cell. To respect the thermal balance, obvious in the steady state of the system, the excessive heat should be fully evacuated by cold water. We check this condition using the model of the cooling structure presented in work \cite{Book}.

\section{Theory}
\subsection{Radiative heat flux}

For any multilayer structure (perhaps, anisotropic and periodically structured in the transverse plane), the RHT (total power flux) into the n-th layer from all other ones can be written as follows (see e.g. in \cite{Circuit Model}):
\e
S_{n}=\displaystyle\frac{1}{2\pi}\sum_{i=1}^{n-1}\int_0^{\infty}{d\omega}\int_0^{\infty}{q\,dq\,P^{i\rightarrow{n}}}.
\label{eq:RHF}
\f
Here, $q$ is the spatial frequency (tangential wave number of the spatial harmonic), and $P^{i\rightarrow{n}}$ represents the double spectral density of power transferred p.u.a. from the i-th layer to the n-th one. For the TM-polarized waves (p-waves) the wave impedance defined as a ratio of the transverse components of the electric and magnetic fields of the eigenmode in each layer is expressed as
\e
Z^{(m)}=\displaystyle\frac{\beta^{(m)}}{\omega\varepsilon_0\varepsilon^{(m)}}\,\,:\,\,m=1,2...,n,
\label{eq:imp}
\f
where $\varepsilon_0$  and $\varepsilon^{(m)}$ are the free-space permittivity and the relative permittivity of the m-th layer, respectively. In the case of the HMM layer we have to replace the
relative permittivity $\varepsilon$ by the transverse component of the effective permittivity $\varepsilon_\perp$. In Eq.~\ref{eq:imp}, $\beta$ represents the propagation factor which is equal to $\beta=\sqrt{k_0^2\varepsilon_\perp-\varepsilon_\perp q^2/\varepsilon_\parallel}$ for each HMM layer, and $\beta=\sqrt{k_0^2\varepsilon-q^2}$ for an isotropic layer. Then, a set of effective resistances ($R_{\rm{eff}}$) is calculated representing the contribution of
layers located above the n-th layer. These resistances are given by:
\e
\begin{split}
&R_{\rm{eff}}^{(n-1)}=R_{\rm{th}}^{(n-1)},\\
&R_{\rm{eff}}^{(n-2)}=F^{(n-1)}R_{\rm{th}}^{(n-2)},\\
&R_{\rm{eff}}^{(n-3)}=F^{(n-1)}F^{(n-2)}R_{\rm{th}}^{(n-3)},\\
&...
\end{split}
\f
where it is denoted:
\e
\begin{split}
&R_{\rm{th}}^{(1)}={\rm{Re}}\left[Z^{(1)}\right],\\
&R_{\rm{th}}^{(m)}={\rm{Re}}\left[Z_{\rm{in-}}^{(m)}\right]-F^{(m)}{\rm{Re}}\left[Z_{\rm{in-}}^{(m-1)}\right]\,\,:\,\,m=2,3,...,n-1,
\end{split}
\f
\e
\begin{split}
&F^{(m)}=\displaystyle\frac{\vert Z^{(m)}\vert ^2}{\vert Z^{(m)}\cos\left(\beta^{(m)}d^{(m)}\right)+j Z_{\rm{in-}}^{(m-1)}\sin\left(\beta^{(m)}d^{(m)}\right)\vert ^2},\\
&Z_{\rm{in-}}^{(m)}=Z^{(m)}\displaystyle\frac{Z_{\rm{in-}}^{(m-1)}+jZ^{(m)}\tan\left(\beta^{(m)}d^{(m)}\right)}{Z^{(m)}+jZ_{\rm{in-}}^{(m-1)}\tan\left(\beta^{(m)}d^{(m)}\right)}.
\end{split}
\f
In the present work the first ($i=1$) and last ($i=n$) layers are assumed to be half-spaces. The double spectral density of transferred power p.u.a. is as follows:
\e
P^{i\rightarrow{n}}=\displaystyle\frac{2}{\pi}\displaystyle\frac{\Theta\left(\omega, T^{(i)}\right)R_{\rm{eff}}^{(i)}}{\vert Z_{\rm{in-}}^{(n-1)}+Z^{(n)}\vert ^2}{\rm{Re}}\left[Z^{(n)}\right]\,\,:\,\,i=1,2,...,n-1.
\label{eq:power}
\f
The function $\Theta\left(\omega,T\right)$ is called Planck's mean energy of a harmonic oscillator, and it is given by
\e
\Theta\left(\omega,T\right)=\displaystyle\frac{\hbar\omega}{\exp{\left(\displaystyle\frac{\hbar\omega}{K_{\rm{B}}T}\right)}-1}.
\label{eq:theta}
\f
Here, $K_{\rm{B}}=1.38\times10^{-23}\,{\rm{J}}/{\rm{K}}$ is the Boltzmann constant, and $\hbar=6.626\times 10^{-34}/2\pi\,{\rm{J}}\cdot{\rm{s}}$.

In the next sections we compare the frequency spectrum of RHT in our structures with that between two flat black bodies. The frequency spectrum of RHT from a black body of temperature $T^{(1)}$ to an adjacent black body does not depend on the thickness of the vacuum gap between them and equals to (see e.g. in \cite{Siegel}):
\e
\displaystyle\frac{dS}{d\omega}=\displaystyle\frac{\omega ^2\mu_0\varepsilon_0}{4\pi ^2}\Theta\left(\omega,T^{(1)}\right).
\label{eq:black body}
\f
This RHT takes into account both TM- and TE-waves, whereas calculating the RHT in our structures we neglect the contribution of TE-waves. Anyway, that of TM-waves is enough to offer the SP RHT in the operational band.

\subsection{Effective-medium model}
For a HMM implemented as a wire medium, two effective-medium models (EMMs) are known: the QEMM and the nonlocal model (see e.g. in the overview \cite{wire_meta}). Both of them describe the optical properties of these media through a uniaxial dyad of effective permittivity:
\e
\overline{\overline{\epsilon}}=\varepsilon_\perp\left(\mathbf{x}_0\mathbf{x}_0+\mathbf{y}_0\mathbf{y}_0\right)+\varepsilon_\parallel\mathbf{z}_0\mathbf{z}_0,
\f
in which $\varepsilon_\perp$ and $\varepsilon_\parallel$ are the transverse and axial components of the dyadic tensor, respectively. According to the QEMM, these components can be expressed as
\e
\begin{split}
&\varepsilon_\perp=\varepsilon_{\rm{h}}\displaystyle\frac{(1+f_{\rm{v}})\varepsilon_{\rm{m}}+(1-f_{\rm{v}})\varepsilon_{\rm{h}}}{(1-f_{\rm{v}})\varepsilon_{\rm{m}}+(1+f_{\rm{v}})\varepsilon_{\rm{h}}},\\
&\varepsilon_\parallel=f_{\rm{v}}\varepsilon_{\rm{m}}+(1-f_{\rm{v}})\varepsilon_{\rm{h}},
\end{split}
\f
where $\varepsilon_{\rm{h}}$ and $\varepsilon_{\rm{m}}$ are the relative permittivities of the host medium and the metal, respectively, and $f_{\rm{v}}={\pi r_0^2}/{a^2}$ ($r_0$ is the wire radius and $a$ is the array period) represents the metal fraction. These simple relations allow to avoid additional boundary conditions required in a more strict non-local model.

\subsection{Electric output and thermal balance}
The overall PV efficiency can be expressed as follows (see e.g. in \cite{Bauer}):
\e
\eta_{\rm{PV}}=\eta_{\rm{OC}}\,.\,\eta_{\rm{QE}}\,.\,\eta_{\rm{FF}}\,.\,\eta_{\rm{UE}}.
\label{eq:efficiency}
\f
Three first efficiency factors depend on the PV semiconductor and are given above. In Eq.~\ref{eq:efficiency}, $\eta_{\rm{UE}}$ is the ultimate efficiency. It is related to the transferred power spectrum $dS/d\omega$  and shows the matching between the absorbed radiation and the PV operational band. This efficiency can be written as follows \cite{Bauer}:
\e
\eta_{\rm{UE}}(\omega_{\rm{g}},\,\omega^-,\,\omega^+)=\displaystyle\frac{\hbar\omega_{\rm{g}}\,.\,Q}{I},
\label{eq:eta}
\f
where $Q$ is the number of photons p.u.a. whose energy is larger than the bandgap one, and $I$ represents the power density for the effective radiation band $(\omega^- \omega^+)$:
\e
Q=\displaystyle\int_{\omega_{\rm{g}}}^{\omega^+}{\displaystyle\frac{\left(\displaystyle\frac{dS}{d\omega}\right)}{\hbar\omega}\,d\omega},\quad
I=\displaystyle\int_{\omega^-}^{\omega^+}{\left(\displaystyle\frac{dS}{d\omega}\right)\,d\omega}.
\label{eq:QI}
\f
If the PV operational band is wider than the range $(\omega_g\,\omega^+)$, the electric output p.u.a. is found as
\e
S_{\rm el.}=\eta_{\rm{PV}}\displaystyle\int_{\omega_{\rm{g}}}^{\omega^+}{{\left(\displaystyle\frac{dS}{d\omega}\right)}\,d\omega}.
\label{output}\f

The total wasted power can be expressed as the difference between the RHT and the electric power p.u.a.:
$$
S_{\rm{w.}}=\int_{\omega^-}^{\omega_{\rm{g}}}{\left(\displaystyle\frac{dS}{d\omega}\right)\,d\omega}+(1-\eta_{\rm{PV}})
$$
\e
\int_{\omega_{\rm{g}}}^{\omega^+}{\left(\displaystyle\frac{dS}{d\omega}\right)\,d\omega}.
\label{eq:wasteP}
\f

For our cooling system shown in Fig.~\ref{TPV2}, the heat power evacuated p.u.a. is calculated through a series connection of several thermal resistances \cite{Book}. For tap water
whose temperature is not lower than 7--8$^{\circ}$C and speed not higher than 4--5 m/s all other resistances but that of the metal wall of thickness $t$  separating the water flow from the PV cell
are negligibly small. Then in accordance to \cite{Book} the evacuated heat p.u.a. is calculated as follows:
\e
S_{\rm{ev.}}=\displaystyle\frac{\lambda_{\rm{w}}}{t}F\Delta T_{\rm{LMTD}},
\label{ev}
\f
where $\lambda_{\rm{w}}=230\,{\rm{Wm^{-1}K^{-1}}}$ is the aluminium thermal conductivity, $F$ is a value of the order of unity (called empirical temperature factor and depending on the hole diameter $D$ and the period of holes), and $\Delta T_{\rm{LMTD}}$ is so-called log-mean temperature difference, defined as \cite{Book}:
\e
\Delta T_{\rm{LMTD}}=\displaystyle\frac{\Delta T_{\rm{in}}-\Delta T_{\rm{out}}}{\ln\left(\displaystyle\frac{\Delta T_{\rm{in}}}{\Delta T_{\rm{out}}}\right)}.
\f

Assuming $T_{\rm{in}}=8^{\circ}$C for the temperature of water in the inlet (typical tap water temperature),
$T_{\rm{out}}=20^{\circ}$C
for that in the outlet (reasonable assumption since the working temperature of the PV cell equals to $25^{\circ}$C), we obtain
$\Delta T_{\rm{in}}=17^{\circ}$ and $\Delta T_{\rm{out}}=5^{\circ}$.
Requirements of mechanical robustness restrict the minimal allowed thickness $t$ of bulk aluminium separating the PV cell from the cold water flow: $t=3-4$ mm \cite{Book}.
The empirical temperature factor of the system in the case when $t$ and $D$ are values of the same orders is $F=0.5$ \cite{Book}. Then the maximal value of heat which can be evacuated corresponds to $t=3$ mm and equals $S_{\rm{ev.\, max}}=38.3\,{\rm{W/cm^{2}}}$. This is the heat evacuated by water heated from 8 to 20$^{\circ}$C.
If the wasted power calculated in Eq.~\ref{eq:wasteP} turns out to be smaller than $38.3\,{\rm{W/cm^2}}$
we may keep $t=3$ mm in (\ref{ev}) decreasing the output temperature of water $T_{\rm{out}}$ so that the thermal balance condition $S_{\rm{ev.}}=S_{\rm w.}$ is still respected. Since the outlet water is heated to lower temperature than 20$^{\circ}$C the regime $S_{\rm w.}<S_{\rm{ev.\, max}}$ means the reliable water cooling.

In the opposite case, when the dissipated radiation p.u.a. noticeably exceeds $38\,{\rm{W/cm^2}}$ it is not possible to respect the thermal balance condition. At a first glance   $S_{\rm{ev.}}=S_{\rm w.}$ can be satisfied raising the value $\Delta T_{\rm{LMTD}}$. However, $T_{\rm out}$ cannot exceed $25^{\circ}$C (practically $\Delta T_{\rm out}$ cannot be smaller than 2--3$^{\circ}$K \cite{Book}). The only way to achieve the thermal balance is to increase $\Delta T_{\rm{in}}$. For fixed tap water temperature it implies the raise of the working temperature of our PV cell. However, for higher temperatures the claimed values of the efficiency and electric output should be reduced in favor of the wasted power. This increase of the wasted power will not allow the thermal balance again. Therefore, the allowed level for the wasted power for $t=3$ mm and tap water is $S_{\rm{w.\, max}}=38\,{\rm{W/cm^{2}}}$. The only opportunity to overcome this limit is to use a more complicated cooling system.

\section{Radiative heat transfer and photovoltaic efficiency}

The temperature of the SiC emitter was assumed to be $T^{(1)}=2000^{\circ}$K. The complex relative permittivity of SiC at this temperature was taken from \cite{SiC}. Optical constants of W for both $T^{(1)}=2000^{\circ}$K (hot nanowires) and for room temperatures (cold nanowires in the second structure) are given in \cite{W}.

\subsection{Nanowires on the hot side}

In the case illustrated by Fig. \ref{fig:d}(a) the PV medium has number $n=6$, and the spectrum of RHT $dS_{6}$/$d\omega$ results from 5 layers located above the plane $y=0$, albeit that the heat produced by layers 4 and 5 is null because both vacuum and bonding film medium are lossless, and $R_{\rm{eff}}^{(4)}=R_{\rm{eff}}^{(5)}=0$. The temperature of layer 2 (first HMM layer) due to its contact to the emitter is assumed to be equal $T^{(2)}=T^{(1)}$. Nanowires in layer 3 (second HMM layer) due to high conductance of tungsten also have the temperature $T^{(1)}$ and $T^{(3)}$ is calculated in accord to their volume fraction. Of course, this model neglects the realistic temperature gradients, however, it is a reasonable model suitable for our qualitative calculations.

In our calculations we compared the contributions of different layers into RHT. It is important to understand the role of HMM layers as additional emitters. Second, it is important to see the impact of nanowires comparing the RHT in our structure with that in the similar structure without nanowires. We also estimate the gain in RHT compared to that between black bodies (\ref{eq:black body}). Finally, we perform a numeric optimization of the structure on following restrictions: $d_4\ge 10$ nm (assuming that the length deviation
$\delta$ is equal to 10 nm), $\va_{2,5}\le 5$ (realistic low-loss media in the NIR range), $d_3+d_4\ge 500$ nm (minimal allowed distance between two parallel surfaces -- a hot and a cold ones -- offered by spacers of micron-gap TPV systems \cite{M2}), and $f_{\rm{v}}<0.3$ (limit of the QEMM validity). Optimization corresponds to the maximal ratio of the useful RHT averaged over the desired range $(\omega_g\,\omega_+)$ to the harmful RHT averaged over the frequency axis beyond this range. There are 7 parameters that we optimized: the fraction $f_{\rm{v}}$ of W in media 2 and 3, the relative permittivities $\va_{2,5}$ of media 2 and 5, and all the thicknesses $d_{2\dots 5}$. Table~\ref{table:1structure} shows the optimal values for these parameters.
\begin{table}[!h]
\centering
\begin{tabular}{|p{1cm}|p{1cm}|p{1.2cm}|p{1.2cm}|p{1cm}|p{1cm}|p{1cm}|}
\hline
{$\varepsilon_2$} & {$\varepsilon_5$} & {$d_2$} & {$d_3$} & {$d_4$} & {$d_5$} & {$f_{\rm{v}}$}\\
\hline
$5$ \ \  \ \ & $5$ \ \ \ \ & $100\,{\rm{nm}}$ \ \ \ \ \ & $600\,{\rm{nm}}$ \ \ \ \ \ & $10\,{\rm{nm}}$ \ \ \ \ & $10\,{\rm{nm}}$ \ \ \ \ & $0.196$\\
\hline
\end{tabular}
\caption{Optimal design parameters for our first structure.}
\label{table:1structure}
\end{table}

\begin{figure*}[!t]
    \centering
    \subfigure[$$]
    {
        \includegraphics[width=6.4cm]{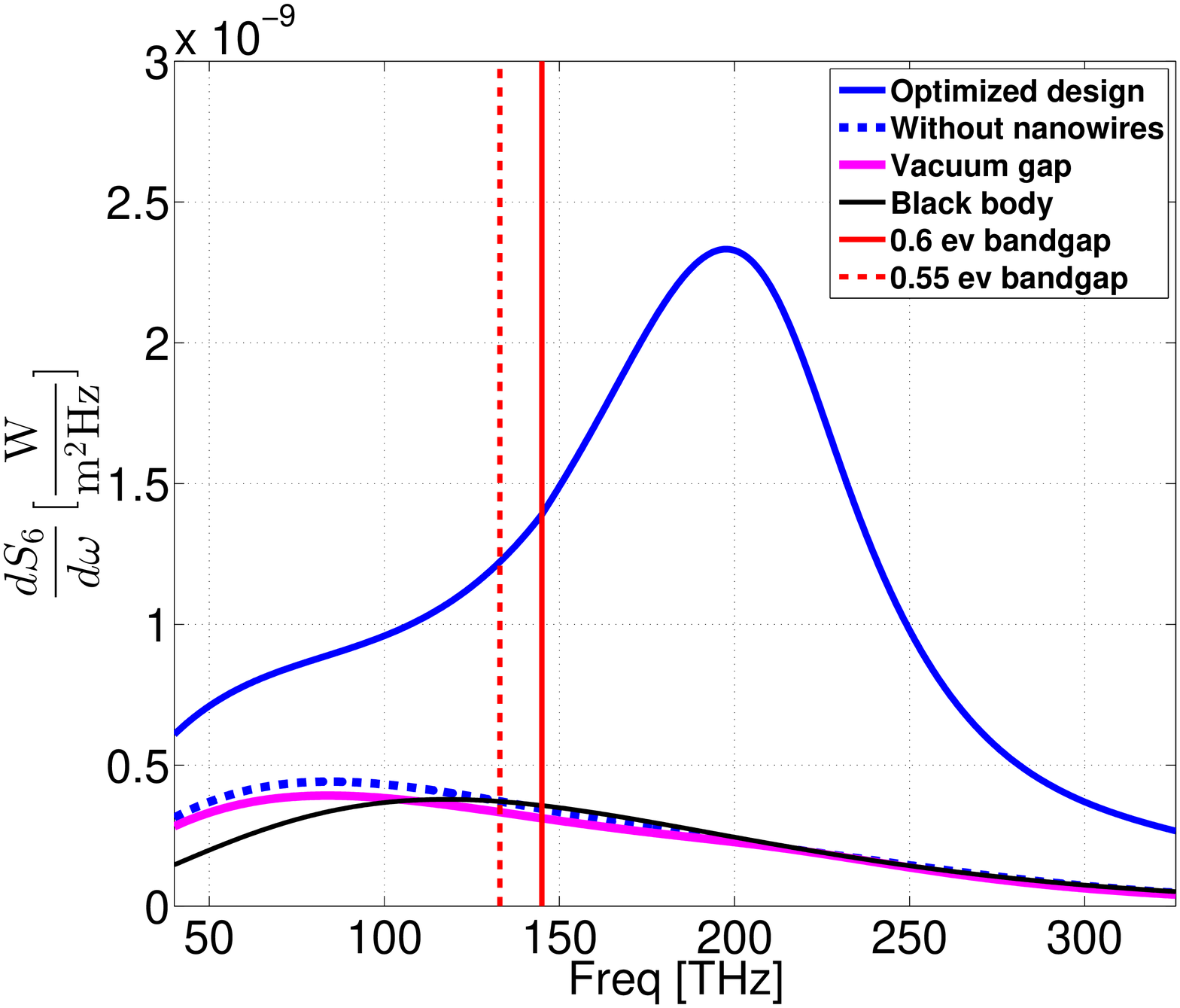}
        \label{fig:dd_1}
    }
    \subfigure[$$]
    {
        \includegraphics[width=6.4cm]{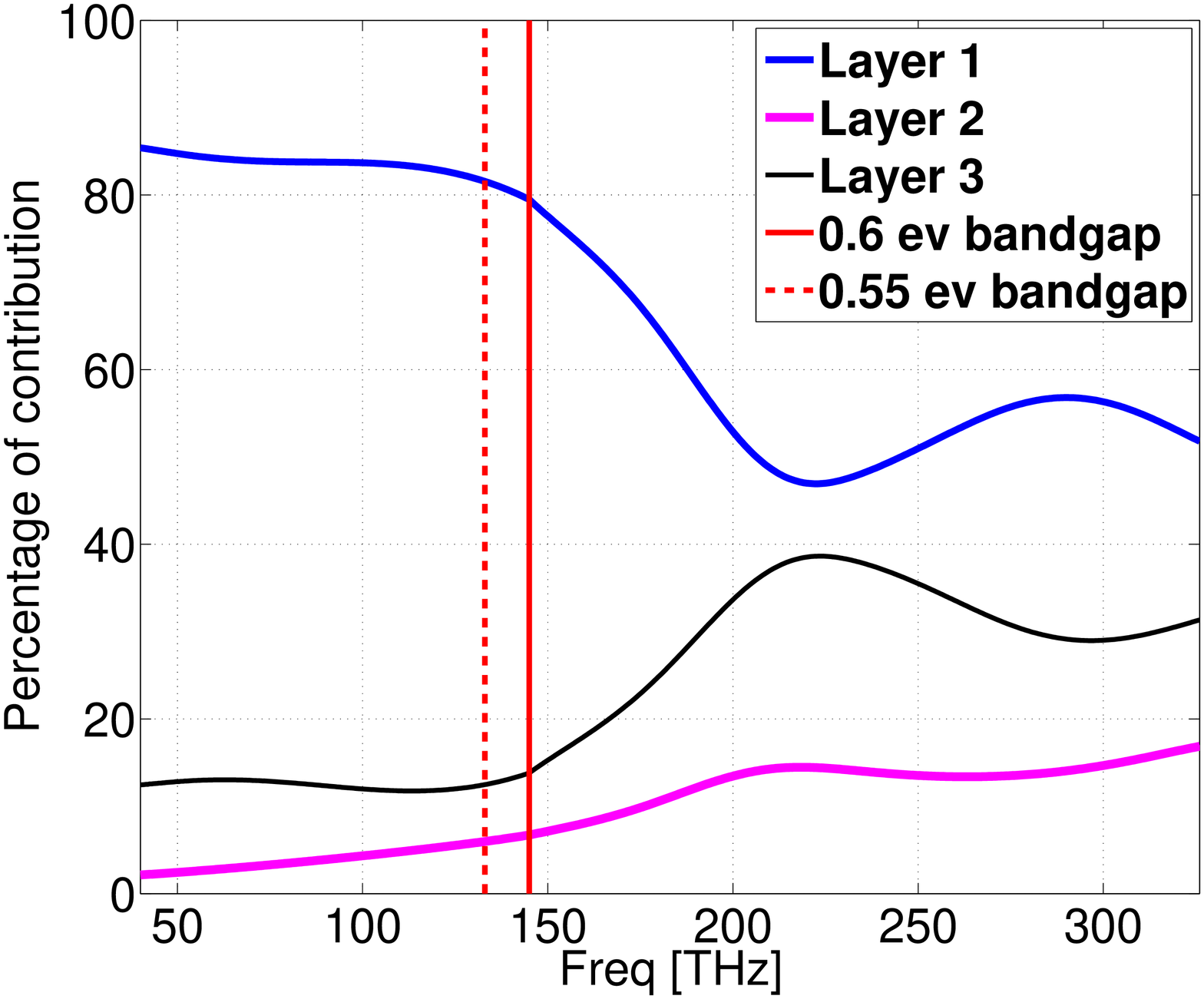}
        \label{fig:dd_2}
   }
    \caption{First structure under study: (a) -- Spectrum of total RHT. The red vertical lines point the bandgap frequency $\nu_b$ for two modifications of the PV material. The PV operational band is between $\nu_b$ and $\nu_+$ (nearly 300 THz).
     (b) -- Percentage of relative contributions of layers 1, 2, and 3 into total RHT spectrum.}
    \label{fig:1RHT}
\end{figure*}

Fig.~\ref{fig:1RHT}(a) shows the spectrum of the total RHT $S_6$ corresponding to these design parameters. It also depicts the RHT spectrum for three reference cases:
a)~the same structure without nanowires, b)~vacuum gap between media 1 and 6, and c)~RHT between two black bodies of same temperatures as that of SiC (2000$^{\circ}$K) and that of InGaAs (25$^{\circ}$C). The operational band is in between $\nu_g=\o_g/2\pi=145$ THz (0.6 eV) or $\nu_g=133$ THz (0.55 eV) and $\nu_+ \approx 300$ THz. Figure~\ref{fig:1RHT}(b) shows the relative contribution of every layer in percent.

By using Eqs.~\ref{eq:eta} and \ref{eq:QI}, we obtain $\eta_{\rm{UE}}=50.1{\%}$. The overall PV  efficiency is approximately $\eta_{\rm PV}=$14.5{\%}.
In the reference case when the nanowires are removed, the ultimate efficiency is close to 32{\%} which results in the overall PV efficiency $\eta_{\rm PV}=$9.3{\%}. Thus, the presence of optimized nanowires not only increases the useful RHT. Due to high ratio of useful RHT to unusable one it noticeably improves the efficiency. Both these factors must result in higher electric output.

\subsection{Nanowires on both sides}

For a structure shown in Fig.~\ref{fig:d}(b)
the temperature of SiC is assumed here to be the same (2000$^{\circ}$K) and $T^{(2),(3)}$ are estimated as above. Again, only three effective layers contribute into RHT $S_7$.

\begin{figure*}[!t]
    \centering
    \subfigure[$$]
    {
        \includegraphics[width=6.4cm]{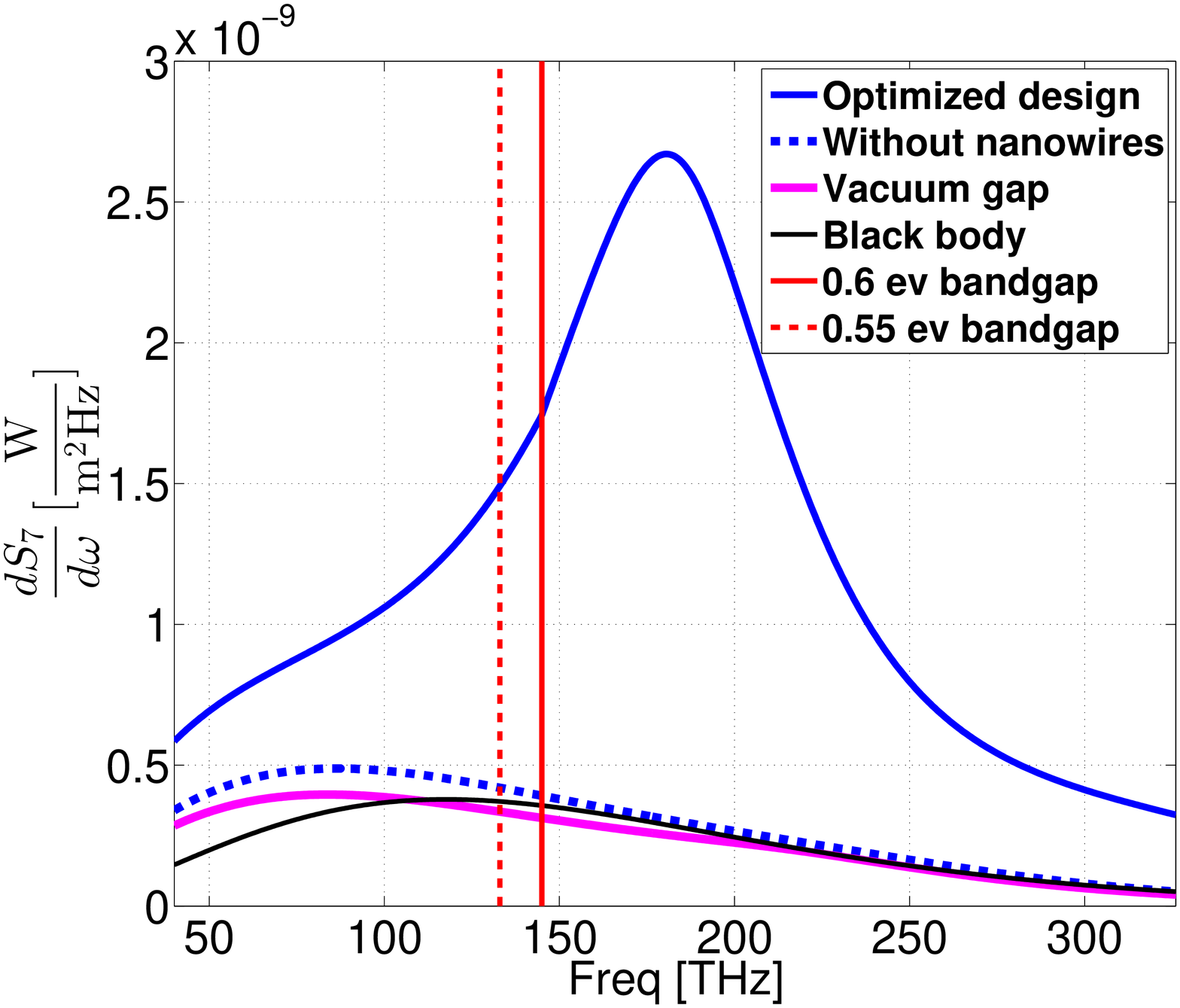}
        \label{fig:ddd_1}
    }
    \subfigure[$$]
    {
        \includegraphics[width=6.4cm]{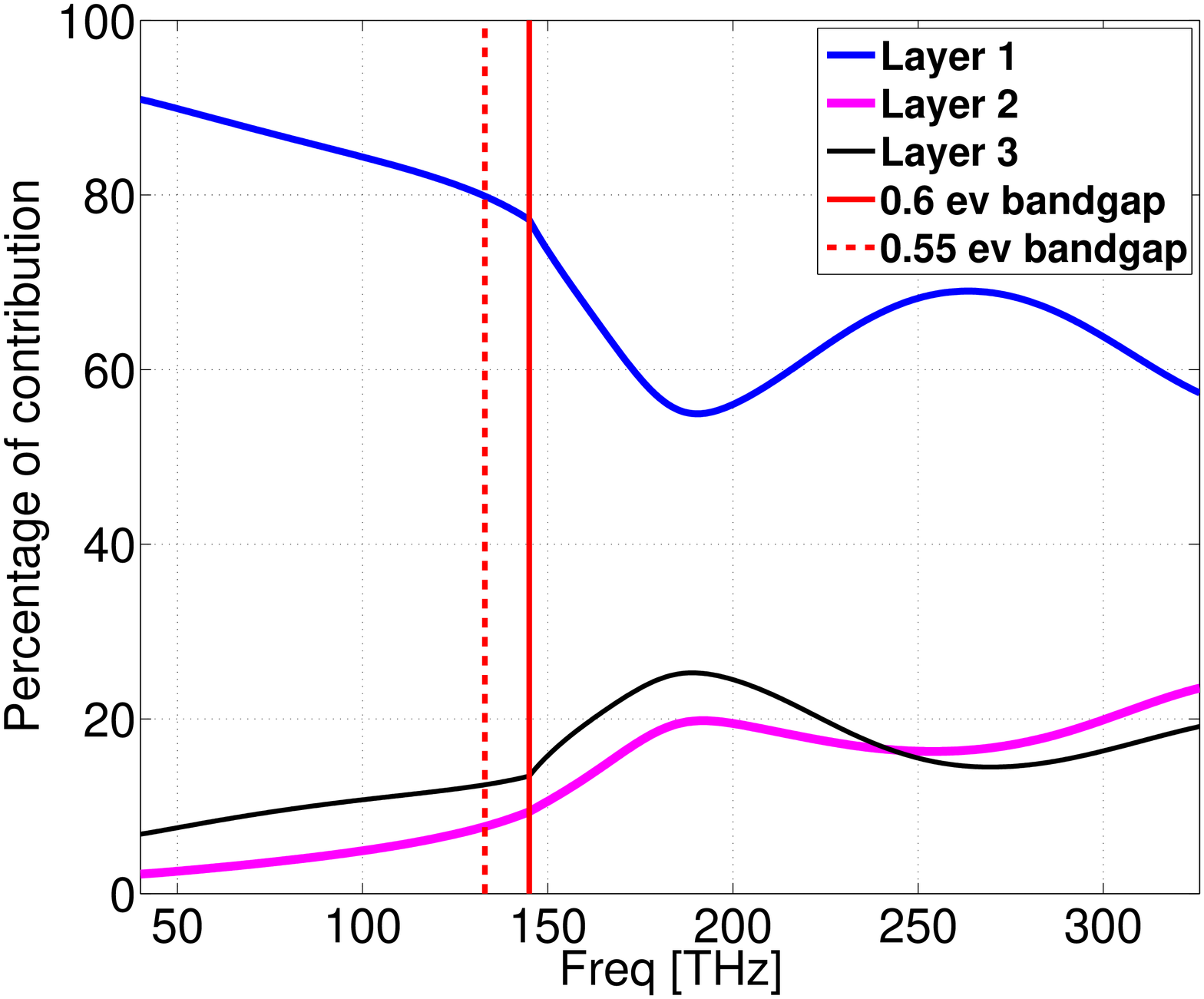}
        \label{fig:ddd_2}
   }
    \caption{Second structure under study: (a) -- Spectrum of total RHT. (b) -- Percentage of relative contributions of layers 1, 2, and 3.}
    \label{fig:2RHT}
\end{figure*}

Though 5-th and 6-th layers are lossy, they do not contribute because they are thermally connected to the PV layer and their temperatures are assumed to be equal $T^{(7)}=25^{\circ}$C. Similar calculations as in the previous subsection have been done for this case. They result in Table \ref{table:2structure} and Fig.~\ref{fig:2RHT}.

\begin{table}[!h]
\centering
\begin{tabular}{|p{0.6cm}|p{0.6cm}|p{1.2cm}|p{1.2cm}|p{1cm}|p{1.2cm}|p{1.2cm}|p{1cm}|}
\hline
{$\varepsilon_2$} & {$\varepsilon_6$} & {$d_2$} & {$d_3$} & {$d_4$} & {$d_5$} & {$d_6$} & {$f_{\rm{v}}$}\\
\hline
$5$ \ \  \ \ & $2$ \ \ \ \ & $100\,{\rm{nm}}$ \ \ \ \ \ & $300\,{\rm{nm}}$ \ \ \ \ \ & $10\,{\rm{nm}}$ \ \ \ \ & $190\,{\rm{nm}}$ \ \ \ \ & $110\,{\rm{nm}}$ \ \ \ \ & $0.196$\\
\hline
\end{tabular}
\caption{Optimal design parameters for our second structure.}
\label{table:2structure}
\end{table}

Here Eqs.~\ref{eq:eta} and \ref{eq:QI} result in $\eta_{\rm{UE}}=50.2{\%}$.
The PV efficiency is again equal $\eta_{\rm PV}$=14.5{\%} in presence of nanowires, and in their absence $\eta_{\rm PV}=$9.3{\%}.

\subsection{Discussion}

So, both suggested structures depicted in Fig. \ref{fig:d} possess the same PV efficiency and nearly same RHT. The same structure without nanowires practically mimics the RHT between two black bodies, and the structure with micron vacuum gap is only slightly worse. We have expected that the optimized second structure with both hot and cold nanowires should be advantageous. However, it is not so.

The higher frequency selectivity for the second structure was expected because it contained one more effective layer. Really, RHT in this case has higher maximum (2.7 $\rm W/m^2Hz$ at the thickness resonance 180 THz versus 2.35 $\rm W/m^2Hz$ at 200 THz for the first structure). However, the unusable RHT is also slightly higher, and the frequency selectivity turns out to be the same. For the first structure the contribution of the free-standing nanowires approaches to that of the emitter at the resonance, though does not reach it. For the second structure the contribution of the SiC emitter strongly dominates, the emission of hot nanowires is not important. However, it is still difficult to make the choice in favor of the first structure, since its optimal design  parameters are quite challenging. Really, the optimal free-standing part of nanowires is much longer than the hosted one, whereas for the second structure this difference is not so great. However, aligned nanowires on both hot and cold sides of the structure are also challenging for nanofabrication. Notice, that small ($\pm$20\%) deviations of design parameters around the optimal ones do not produce great changes in our results, but even with these allowed deviations and allowed aperiodic arrangement of nanowires both suggested structures are still difficult to manufacture (though simpler than that suggested in \cite{OE}). This difficulty is an additional reason to replace silicon carbide by tungsten. In this case hot W nanowires can be prepared as protrusions on the W substrate.

\subsection{Metal emitter}

\begin{figure}[htb!]
\begin{center}
\subfigure[1st structure]{\includegraphics[width=6cm]{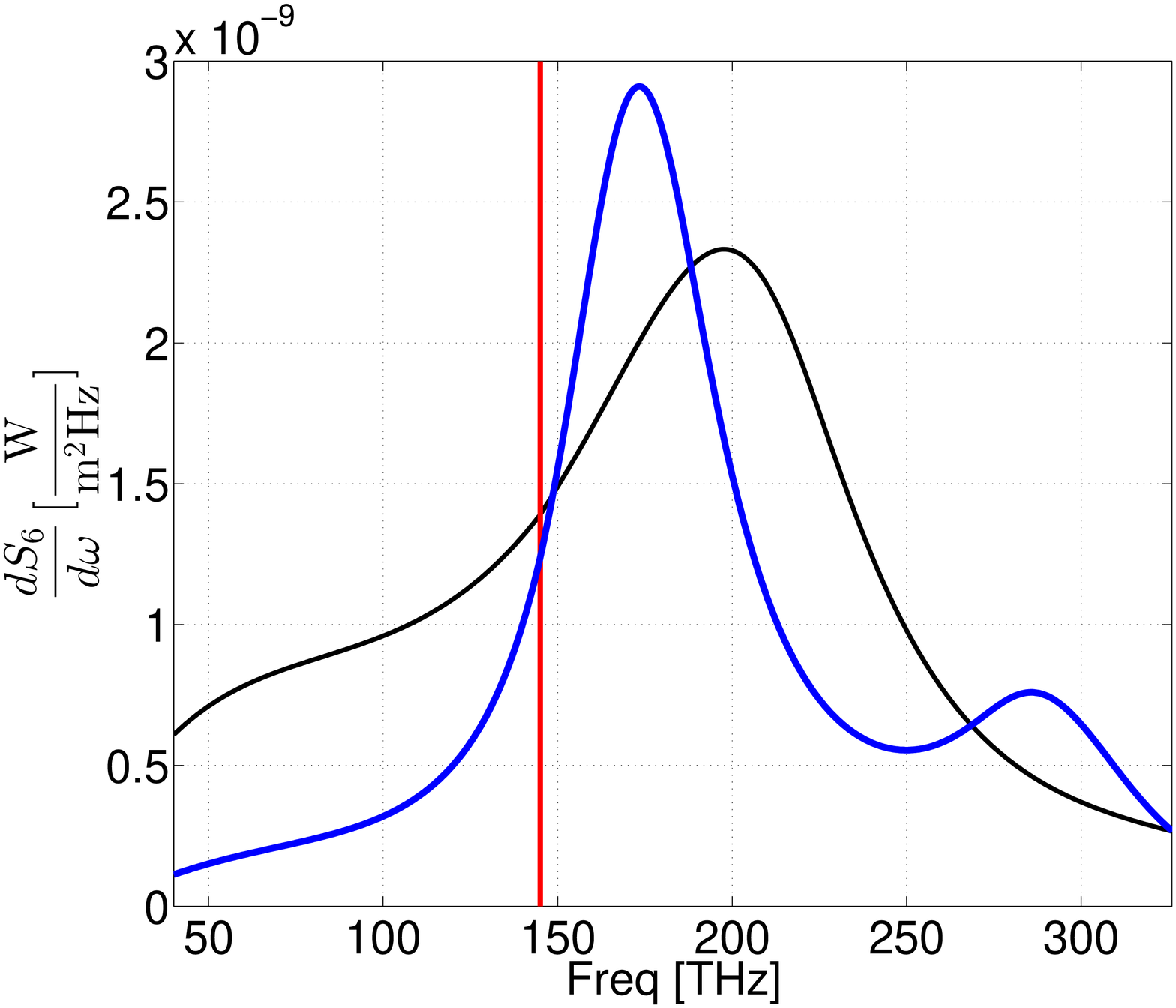}}
\subfigure[2nd structure]{\includegraphics[width=6cm]{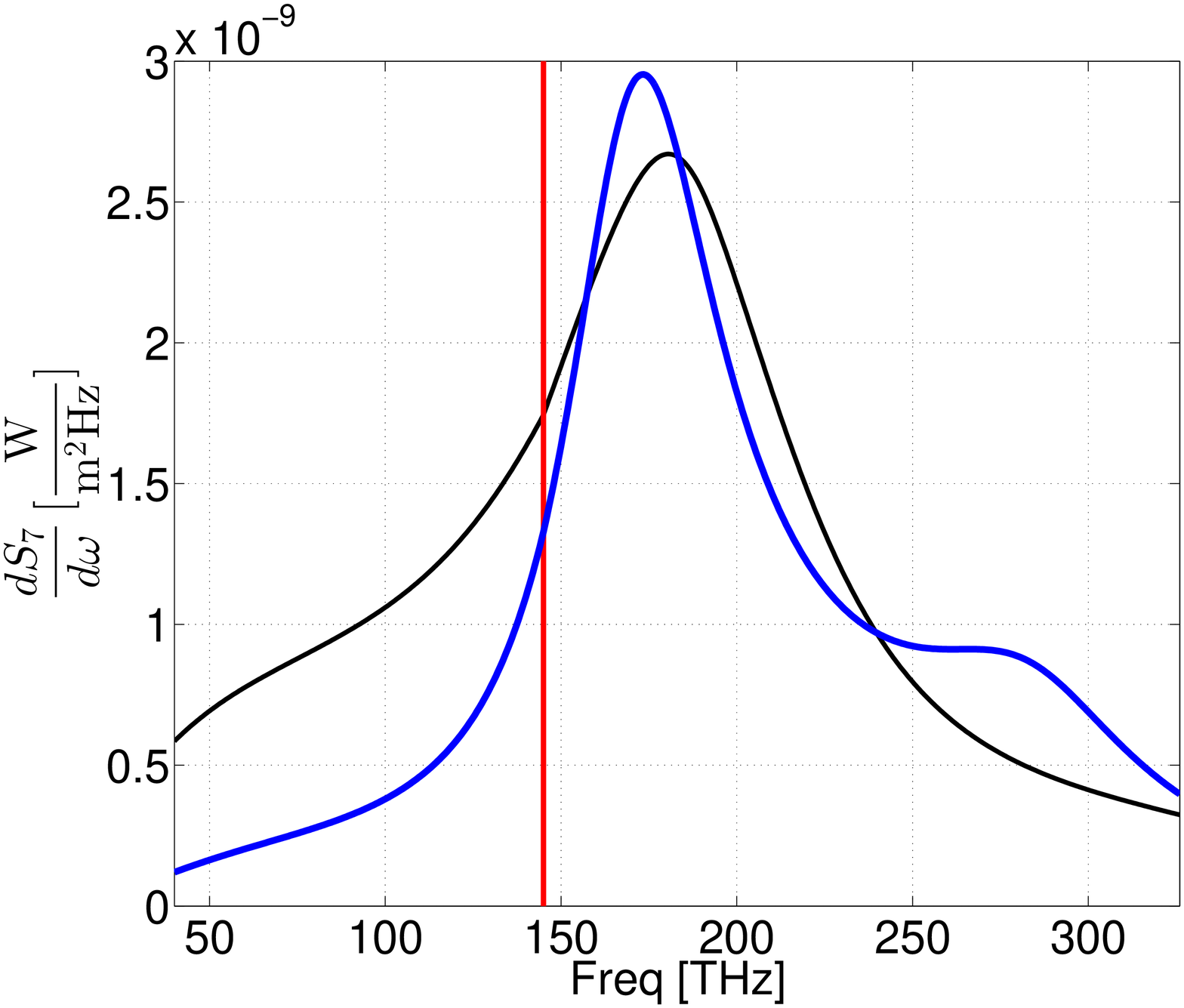}}
\caption{Spectrum of total radiative heat flux. Black and blue Solid curves correspond to SiC and W emitters, respectively. (a) -- First structure. (b) -- Second structure.}
\label{fig:RHT_metal}
\end{center}
\end{figure}

Figure~\ref{fig:RHT_metal} compares the spectrum of total RHT for two cases: SiC emitter and W emitter. For the first structure the replacement of SiC by W has not resulted in the change of the optimal design parameters: all of them are same as in Table~\ref{table:1structure}. The optimized values for the second structure are presented in Table~\ref{table:2structure_metal}.

\begin{table}[!h]
\centering
\begin{tabular}{|p{0.6cm}|p{0.6cm}|p{1.2cm}|p{1.2cm}|p{1cm}|p{1cm}|p{1cm}|p{1cm}|}
\hline
{$\varepsilon_2$} & {$\varepsilon_6$} & {$d_2$} & {$d_3$} & {$d_4$} & {$d_5$} & {$d_6$} & {$f_{\rm{v}}$}\\
\hline
$5$ \ \  \ \ & $5$ \ \ \ \ & $120\,{\rm{nm}}$ \ \ \ \ \ & $420\,{\rm{nm}}$ \ \ \ \ \ & $10\,{\rm{nm}}$ \ \ \ \ & $70\,{\rm{nm}}$ \ \ \ \ & $50\,{\rm{nm}}$ \ \ \ \ & $0.2$\\
\hline
\end{tabular}
\caption{Optimized values for host media permittivities, thicknesses, and metal fraction for the second structure and W emitter.}
\label{table:2structure_metal}
\end{table}

As it can be seen in Fig.~\ref{fig:RHT_metal}, the spectrum of RHT is more narrow-band for the W emitter. The most important feature is higher suppression of low frequencies (below $\nu_b$). This suppression causes the increase in the ultimate efficiency from $\eta_{\rm UE}=$50.1{\%} to $\eta_{\rm UE}=$61.8{\%} and from 50.2{\%} to 60.8{\%} for the first and second structures, respectively. {Therefore, the photovoltaic efficiency equals $\eta_{\rm{PV}}=17.9{\%}$ and $\eta_{\rm{PV}}=17.6{\%}$ for the first and second structures, respectively. The equivalence of efficiencies in this case allows us to choose the first structure, evidently simpler for fabrication, for further investigation in our next papers.

\section{Electric output and thermal balance}

The values of efficiencies obtained in the cases of the SiC emitter in accordance to (\ref{output}) correspond to the following output electric powers:
$S_{\rm{el.}}=3.3\,{\rm{W/cm^{2}}}$ and $S_{\rm{el.}}=3.4\,{\rm{W/cm^{2}}}$ for the first and second structures, respectively. The small difference results from slightly higher useful RHT in the second structure. Dissipated power p.u.a. for the first structure is equal $S_{\rm{w.}}=29.8\,{\rm{W}}/{\rm{cm}}^2$, and for the second structure $S_{\rm{w.}}=31.4\,{\rm{W}}/{\rm{cm}}^2$.
These values are below the limit $38.3\,{\rm{W}}/{\rm{cm}}^2$ for the heat p.u.a. evacuated by the tap water on pre-supposed conditions, i.e. the thermal balance corresponds to the lower outlet temperature than $20^{\circ}$C. For the geometry with interdigital nanowires suggested \cite{OE} the dissipated power would exceed $S_{\rm{w.}}=50\,{\rm{W}}/{\rm{cm}}^2$ (for tungsten nanowires). This power cannot be evacuated by tap water. Therefore, the interdigital arrangement of nanowires is really useless.

In the case of the W emitter we obtain $S_{\rm{el.}}=3.8\,{\rm{W/cm^{2}}}$ and $S_{\rm{el.}}=4.3\,{\rm{W/cm^{2}}}$ for the first and second structures, respectively. This 15\% difference occurs because the useful RHT in the second structure is 15\% higher. However, in the case of the tungsten emitter the low-frequency radiation is better suppressed for the first structure, that corresponds to the wasted power p.u.a. $S_{\rm{w.}}=21.5\,{\rm{W}}/{\rm{cm}}^2$. For the second structure it is equal $S_{\rm{w.}}=25\,{\rm{W}}/{\rm{cm}}^2$. Both these values allow the safe water cooling, however, their comparison is one more reason in favor of the first geometry.

\section{Conclusions}

In this work we have introduced new design solutions for metamaterial-enhanced micron-gap TPV systems
which are much closer to their practical implementation than the similar structures we have proposed earlier.
We have done a system analysis of the operation of our TPV systems. This analysis besides the calculation of radiative heat transferred to the PV panel,
includes also extended calculations of the overall PV efficiency, electric output and dissipated heat.
Extended calculations allowed the optimization of newly suggested structures. We have proved that their operation is compatible with principles of
standard water cooling. The claimed values of the total PV efficiency (14.5--17.9\%) are only slightly better than those achieved for available TPV systems (12--13\%) operating at the same temperatures \cite{Bauer}. However, this enhancement was not our main purpose. The last one was to obtain high values for the electric output p.u.a. We claim 3.3--4.3 ${\rm{W}}/{\rm{cm}}^2$, that 3--4 times exceeds those achieved in the best available TPV systems. This gain results from the enhanced useful RHT whereas the dissipated RHT is safely evacuated by water.

Notice, that the TPV efficiency is an irrelevant parameter for our systems, since the emitter cannot be shared out from the system. The hot half-space in our model is coupled to the metamaterial layers and through them -- to the half-space of semiconductor. Moreover, hot metamaterial layers strongly contribute into RHT. It is possible to characterize the total efficiency of our system in terms of the electric output normalized to the fuel energy spent for heating the unit area of the emitter. However, for the instance, we have not studied this thermal process. In our future works we aim to estimate this effect taking into account the finite thickness of the emitter.

It worth noticing, that checking the thermal balance condition we neglect the temperature gradient over the PV cell, though in the calculation of RHT we model the PV medium by a half-space. There is no contradiction. Even a tiny PV layer of several microns absorbs practically all the incident infrared light and can be replaced by a half-space in electromagnetic simulations. The rear electrode of the PV cell as a rule is performed as a solid metal film. So,  the thermal conductance across the whole PV cell is sufficiently high to assume its bottom surface on the same temperature as that of the top one.

We hope that this work will approach us to the creation of an advantageous TPV system enhanced by metal nanowires operating as a very efficient electric generator.

\section{Glossary}

TPV -- thermophotovoltaic; PV -- photovoltaic; BW -- bandwidth; HMM -- hyperbolic metamaterial; EMM -- effective-medium model;
QEMM -- quasi-static effective-medium model; RHT -- radiative heat transfer; SP -- super-Planckian; THz -- terahertz; p.u.a. -- per unit area.


\end{document}